\documentclass[11pt]{article}
\usepackage{moriond,epsfig}

\def\be{\begin{equation}}
\def\ee{\end{equation}}
\def\bea{\begin{eqnarray}}
\def\eea{\end{eqnarray}}

\newcommand{\Photo}{}

\begin{document}

\vspace*{1cm}

\begin{flushright}
Edinburgh 2013/06
\end{flushright}

\vspace*{3cm}

\title{PROGRESS IN THE NNPDF GLOBAL ANALYSIS}

\author{Christopher S. Deans}

\address{On behalf of the NNPDF Collaboration\\Tait Institute, University of Edinburgh,\\
JCMB, KB, Mayfield Rd, Edinburgh EH9 3JZ, Scotland}

\maketitle\abstracts{
We report on recent progress in the NNPDF framework
of global PDF analysis. The NNPDF2.3 set is the first and only
available PDF set with includes LHC data. 
A recent benchmark comparison of NNPDF2.3 and all other modern NNLO PDF
sets with LHC data was performed. We have also studied theoretical
uncertainties due to heavy quark renormalization schemes,
higher twists and deuterium corrections in PDFs.
Finally, we report on the release of positive definite
PDF sets, based on the NNPDF2.3 analysis, specially suited
for use in Monte Carlo event generators.
}

\paragraph{Introduction}

In this age of precision QCD at the LHC, the determination of the parton 
distribution functions which characterize the structure of the proton 
is becoming increasingly important. For some measurements the size of
PDF uncertainties approaches that of experimental uncertainties. However, the LHC also
offers a great opportunity to constrain PDFs. Some LHC observables
important to PDFs have already been released using 2010 and 2011 data,
and many others are now being analysed with the full 2012 data. NNPDF2.3
is the first and only publicly available PDF set which includes LHC 
data~\cite{Ball:2012cx}, extending the dataset from 
NNPDF2.1~\cite{Ball:2011mu,Ball:2011uy} with data on jet and electroweak
boson production from the ATLAS, CMS and LHCb collaborations. The impact 
of the data was found to be moderate, but it will certainly increase
with the release of more data with greater precision and including a 
wider range of different physical observables.

Here we will discuss recent work involving members of the NNPDF collaboration: 
a benchmarking of NNLO PDF sets, investigation of sources of theoretical
uncertainties on PDF determinations, and the release of a NNPDF2.3 set
for NLO Monte Carlo generators. 

\paragraph{PDF benchmarking with LHC data}

\begin{figure}[t]
    \begin{center}
\includegraphics[width=0.48\textwidth]{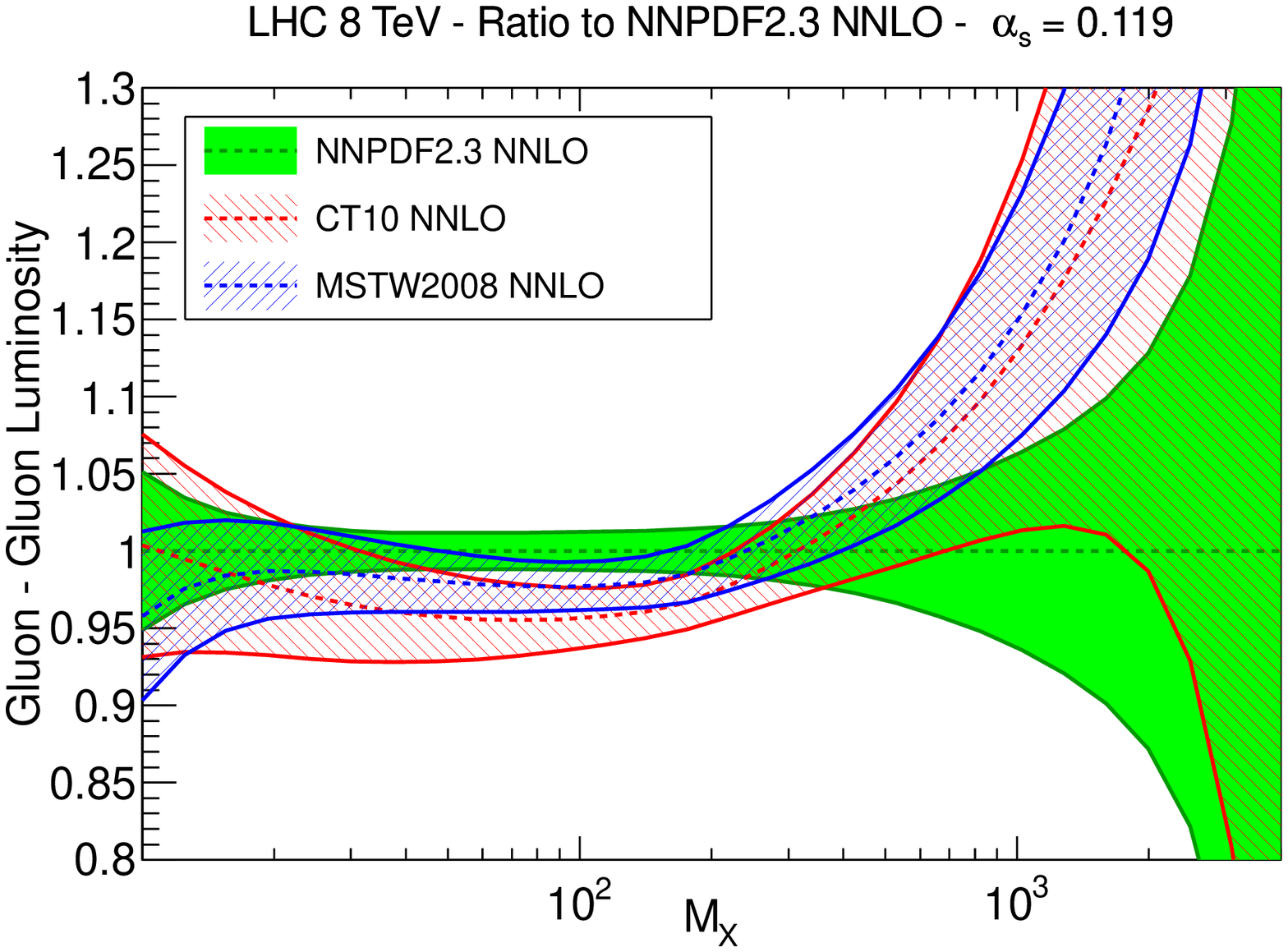}
\includegraphics[width=0.48\textwidth]{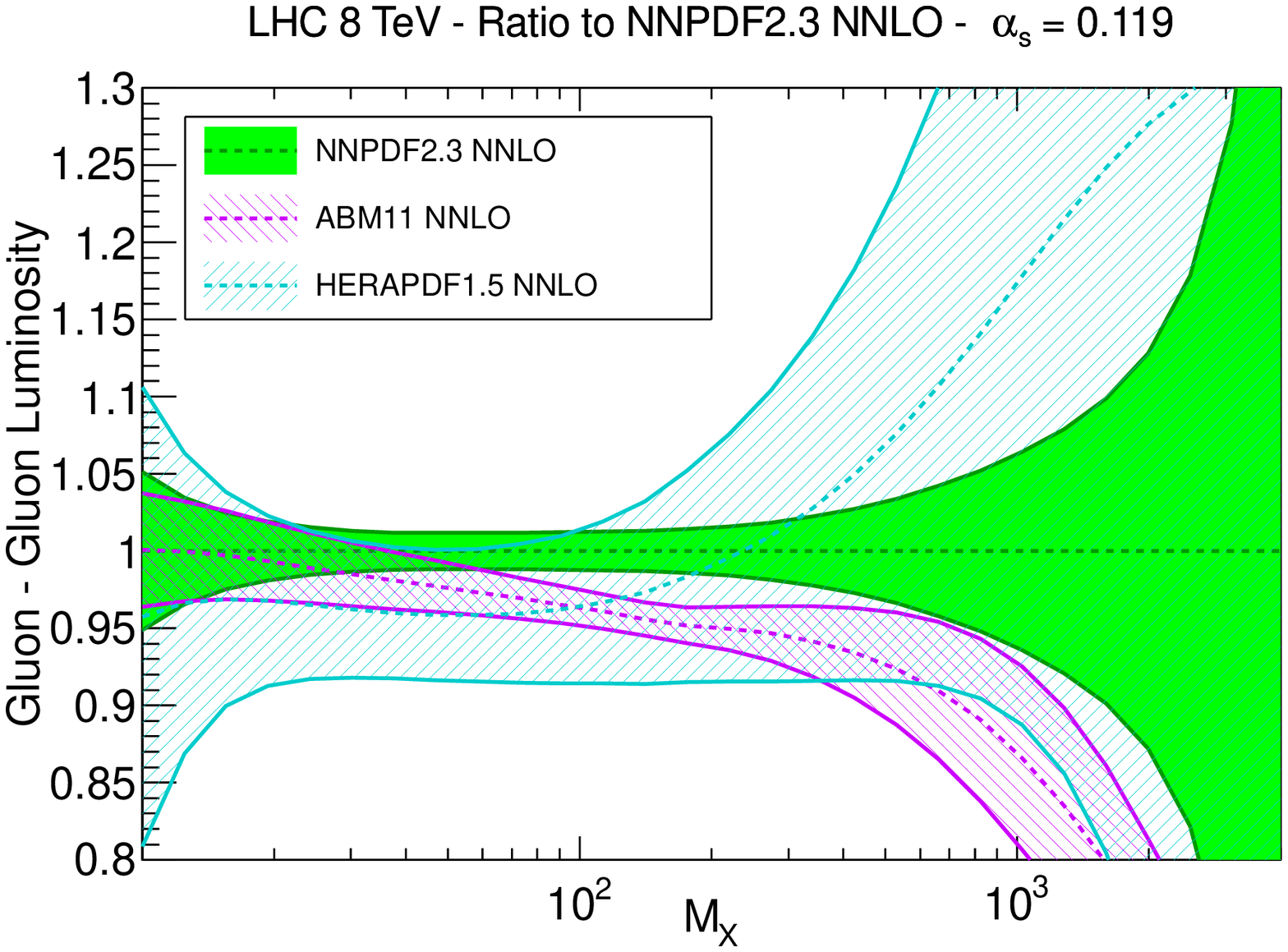}
      \end{center}
     \caption{\small 
    \label{fig:lumi} The gluon--gluon parton
luminosity at $\sqrt{s}=8$~TeV for
$\alpha_s(M_Z)=0.119$ as a function of the invariant mass
of the final state $M_X$, for various NNLO PDF sets.}
\label{gglumi}
\end{figure}

An extensive benchmarking exercise comparing the modern NNLO PDFs was 
carried out in Ref.~\cite{Ball:2012wy}. It benchmarked both PDFs and 
parton luminosities for NNPDF2.3, ABM11, HERAPDF1.5, CT10 and MSTW08,
and also compared the predictions of these sets to LHC data both at the 
level of inclusive cross sections and of differential distributions. The
comparisons to data were made quantitative using a variety of $\chi^2$
estimators. The gluon--gluon luminosities for the PDF sets in the
benchmark are shown as a ratio to NNPDF2.3 in Fig.~\ref{gglumi} for 
$\alpha_S=$ 0.119. 

There is good agreement between the three global PDF
sets (NNPDF2.3, MSTW08 and CT10), which is at least as good at NNLO as
was previously found for NLO. For several LHC processes the combined 
PDF+$\alpha_S$ uncertainties determined from the envelope of MSTW, NNPDF
and CT sets are found to be smaller in the 2012 PDF sets than for the 2010 sets. The HERAPDF1.5
central values also show good agreement with the global sets, with 
larger uncertainties due to the smaller dataset used. For ABM11, 
however, disagreement was found for several PDFs and LHC cross sections.
Possible reasons for this disagreement are the use of an FFN scheme or
the inclusion of higher twist corrections in ABM11, discussed below, but
the situation will become more clear with the release of further precise
LHC measurements, particularly top pair production\cite{Czakon:2013tha}.

\paragraph{Theoretical uncertainties in PDF analyses}

There has also been recent work on several
theoretical issues in PDF determination and associated uncertainties
which may have some impact on current PDF sets and explain some of the
differences between them. In particular, Ref.~\cite{Ball:2013gsa} looked at
the use of fixed-flavor number (FFN) versus
variable-flavor number (VFN) renormalization schemes, higher twist
corrections, and nuclear corrections. The analysis was performed in the
NNPDF2.3 framework by performing fits with different 
theoretical assumptions and comparing the results, both in terms of PDFs
and LHC cross sections.

The treatment of heavy quarks in PDF fits can potentially have a large
impact on the results obtained~\cite{Thorne:2012az}. Most modern PDF 
sets, including NNPDF2.3, use a variable flavor number scheme, where 
fixed order calculations including heavy quark masses are combined with 
all-orders resummation of contributions from perturbative evolution. 
Other sets, for instance ABM11, instead use a 
fixed flavor number scheme where the heavy quarks are not treated as active flavors and do 
not enter into QCD evolution equations. 
The effect of scheme dependence was studies by producing versions of the NNPDF2.3 global pdf fit using the FFN scheme,
rather than the FONLL VFN which is the default~\cite{Forte:2010ta}\footnote{The NNPDF2.3 sets using the FFN scheme are available on the NNPDF hepforge webpage: \url{nnpdf.hepforge.org/}}.

It was found that first of all the differences between FFN and VFN
can be large, and in particular the use of a FFN leads to a suppressed
large-$x$ gluon distribution and increased medium and small-$x$ quark
distributions. The $\chi^2$ fit quality for the two
schemes was also studied, and it was found that FFN provides a systematically poorer
description of the data, especially for the high-$Q^2$
and small-$x$ HERA data, consistent with expectations given that
the FFN scheme does not include the resummation of DGLAP
logarithms in the PDF evolution which are potentially large in this 
kinematic region.

\begin{figure}[h]
    \begin{center}
\includegraphics[width=0.75\textwidth]{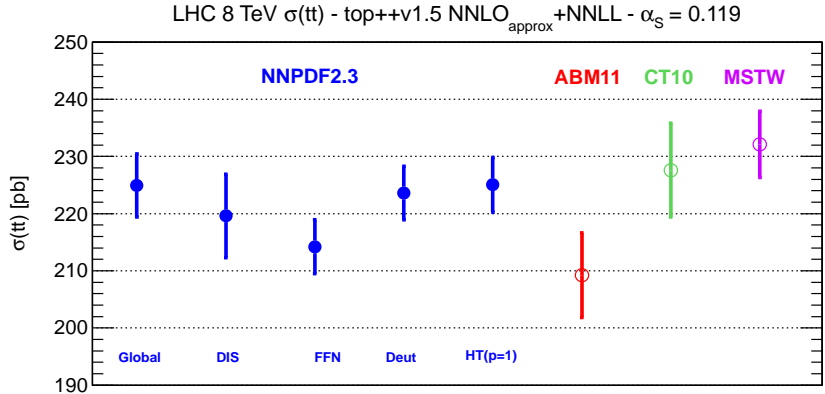}
      \end{center}
     \caption{\small 
    \label{fig:lhcxsec} The $t\bar{t}$ computed
at NNLO$_{\rm approx}$ using
    various PDF sets  at $\sqrt{s}=8$~TeV for
$\alpha_s(M_Z)=0.119$.}
\end{figure}

In Fig.~\ref{fig:lhcxsec} we show the $t\bar{t}$ total cross section 
computed at NNLO$_{\rm approx}$ using {\tt top++} for various PDF sets at 
$\sqrt{s}=8$~TeV for a common value of $\alpha_s(M_Z)=0.119$. Is clear that 
both the use of a reduced DIS-only dataset and of a FFN scheme
bring NNPDF2.3 in better agreement with the ABM11 prediction.

Ref.~\cite{Ball:2013gsa} also looked at the impact of including corrections for higher twist 
and deuterium nuclear effects. Higher twist effects arise from 
power-suppressed contributions to the Wilson expansion, and whilst 
low-scale data most sensitive to these effects are commonly removed by 
kinematic cuts, there may still be some residual effects. An NNPDF2.3 
fit was performed with the leading twist DIS structure functions 
supplemented with a twist four correction taken from Ref.~\cite{Alekhin:2012ig}. It was found that the inclusion 
of this correction had a negligible effect on the fit within 
uncertainties, and even when this size of these corrections were doubled
the impact was marginal. Fits including nuclear 
corrections to structure function data on deuterium targets were also performed. Here the 
corrections do have a moderate impact, however it was limited to the 
large-$x$ up-down separation. The $t\bar{t}$ total cross section for 
fits with higher twist and deuterium corrections are also show in 
Fig.~\ref{fig:lhcxsec}, the differences for this particular observable 
are negligible.

\paragraph{Positive definite PDFs for Monte Carlo generators}

Parton distribution functions are not positive definite
quantities beyond leading order, and thus some PDF groups
allow for negative PDFs in their NLO and NNLO sets.
This is the case in the NNPDF2.3 analysis, where
instead the positivity of several physical cross sections
is imposed as an additional constraint during the minimization. In practice, 
negative PDFs then arise only in phase space regions where experimental
constraints on a given PDF combination are very scarce, such as the
small-$x$ gluon or the large-$x$ antiquarks, and reflect the large
PDF uncertainties in such extrapolation regions.

While there are no theoretical issues in dealing with negative
PDFs in fixed order computations, some issues can arise
in the context of Monte Carlo event generation. For instance, only positive PDFs can be used
in the Sudakov form factors that determine the shower probabilities,
and negative small-$x$ PDFs may affect the description of
the underlying event (UE) and of multiple interactions (MPI) in the event generator.
In addition, there can also be practical issues, such as numerically inefficiencies when reweighting events
starting from negative PDFs.

None of these issues are insurmountable. For instance the {\tt Sherpa}~\cite{Gleisberg:2008ta}
event generator includes tunes for the default NNPDF2.3 set, which
correctly describe UE and MPI at hadron colliders, and
NLO event generators such as {\tt aMC@NLO}~\cite{Frederix:2011ss} provide exact PDF
reweighting without the requirement of positive definite PDFs.
In addition, it is a common practice to use separate PDFs
for the hard matrix element and for the parton shower, since the latter
includes the information from the non-perturbative tunes to the data.

Despite the absence of any conceptual issue, we believe it is
useful for the community to have positive definite
NLO NNPDF sets for use in event generators at the LHC.
To achieve this, we have taken as a starting point the LHAPDF grids for various
NNPDF2.3 NLO sets. In the interpolating grid for each PDF,
any $x,Q$ point which has a negative value is instead assigned a very
small positive number. This defines the
NNPDF2.3 MC PDF sets. It was also necessary to modify the LHAPDF
wrapper of these sets to ensure that the output is positive definite
in any cases (to avoid negative interpolation between grid points).

\begin{figure}[h]
    \begin{center}
\includegraphics[width=0.49\textwidth]{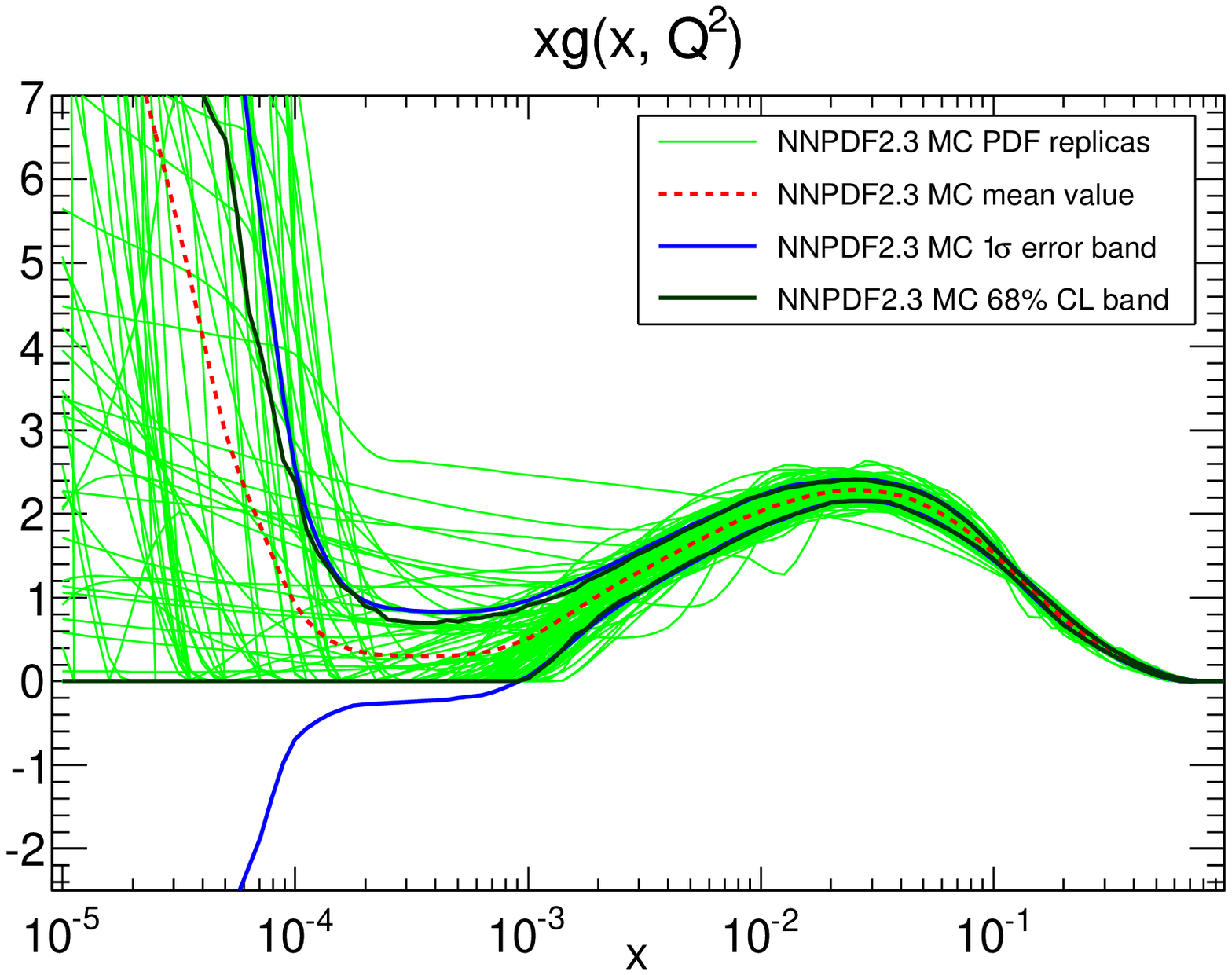}
\includegraphics[width=0.49\textwidth]{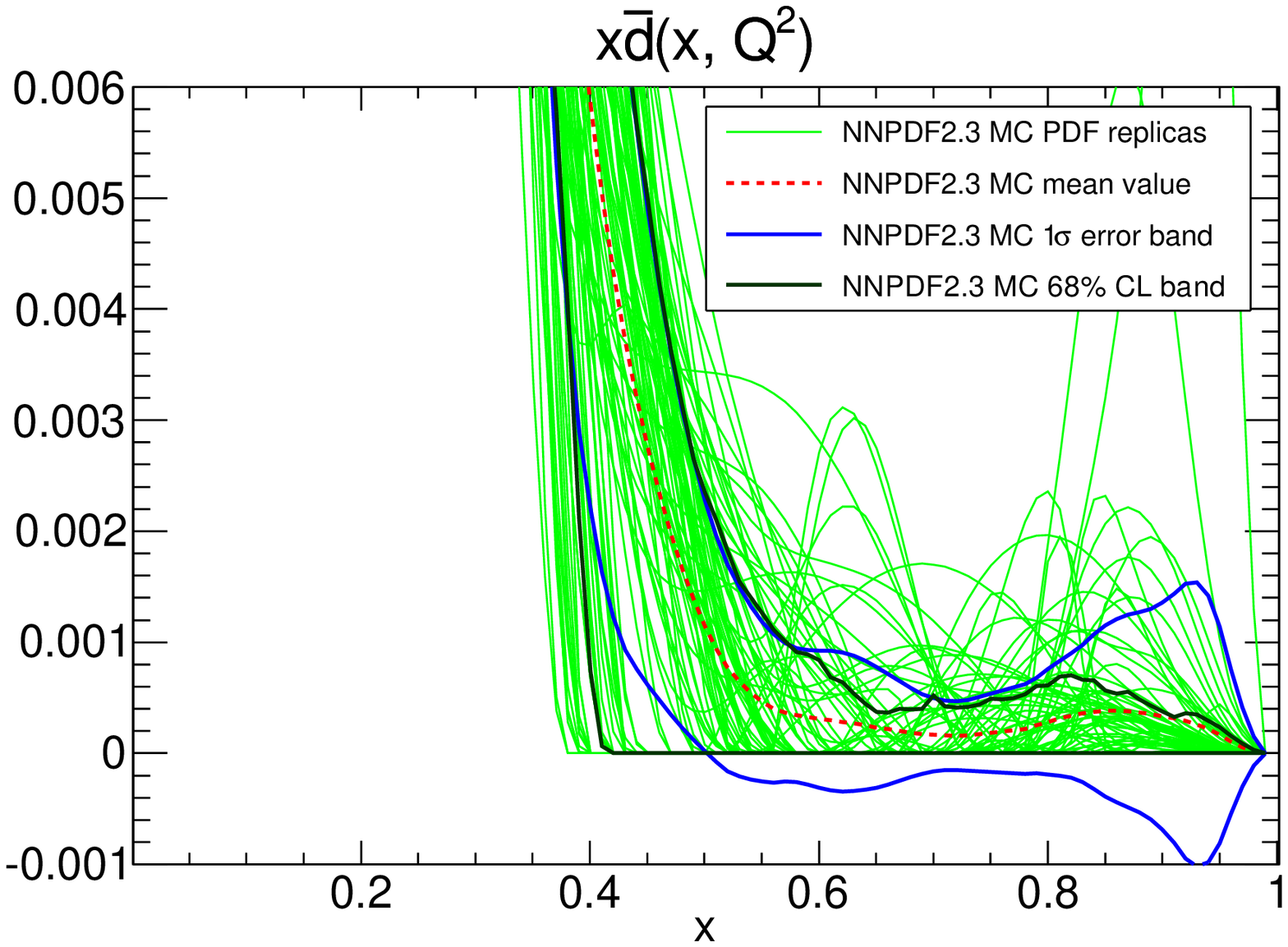}
      \end{center}
     \caption{\small 
    \label{fig:pdfmc} The small-x gluon PDF (left plot)
and the large-x $\bar{d}$ PDF (right plot) at $Q^2=2$ 
GeV$^2$ in the {\tt NNPDF23\_nlo\_as\_0119\_mc.LHgrid}
PDF set. Each of the green lines is one PDF replica.
As can be seen, in these NNPDF2.3 MC sets both the
central values and the individual replicas are
positive definite.}
\end{figure}

Starting from LHAPDF v5.8.9, these Monte Carlo NNPDF2.3 NLO sets
are publicly available. 
In Fig.~\ref{fig:pdfmc} we show small-x gluon PDF (left plot)
and the large-x $\bar{d}$ PDF (right plot), both at $Q^2=2$ 
GeV$^2$, from the {\tt NNPDF23\_nlo\_as\_0119\_mc.LHgrid}
PDF set. We have explicitly checked that the good $\chi^2$ quality description
of all datasets in the global PDF analysis is unaffected, 
which was expected given that the PDFs in the default sets only go negative in regions far
from experimental constraints.

In summary, the NNPDF2.3 NLO MC PDF sets are positive definite
PDF that are specially suited for their use in event generators,
while maintaining the excellent description of all hard scattering
data, and in particular of all relevant LHC data, which 
characterizes the default NNPDF2.3 sets.

\section*{References}
\begin{thebibliography}{99}

\bibitem{Ball:2012cx}
  R.~D.~Ball, V.~Bertone, S.~Carrazza, C.~S.~Deans, L.~Del Debbio, S.~Forte, A.~Guffanti and N.~P.~Hartland {\it et al.},
  Nucl.\ Phys.\ B {\bf 867} (2013) 244
  [arXiv:1207.1303 [hep-ph]].

\bibitem{Ball:2011mu}
  R.~D.~Ball, V.~Bertone, F.~Cerutti, L.~Del Debbio, S.~Forte, A.~Guffanti, J.~I.~Latorre and J.~Rojo {\it et al.},
  Nucl.\ Phys.\ B {\bf 849} (2011) 296
  [arXiv:1101.1300 [hep-ph]].

\bibitem{Ball:2011uy}
  R.~D.~Ball {\it et al.}  [NNPDF Collaboration],
  Nucl.\ Phys.\ B {\bf 855} (2012) 153
  [arXiv:1107.2652 [hep-ph]].

\bibitem{Ball:2012wy}
  R.~D.~Ball, S.~Carrazza, L.~Del Debbio, S.~Forte, J.~Gao, N.~Hartland, J.~Huston and P.~Nadolsky {\it et al.},
  JHEP in press
  [arXiv:1211.5142 [hep-ph]].

\bibitem{Czakon:2013tha}
  M.~Czakon, M.~L.~Mangano, A.~Mitov, and J.~Rojo,
  arXiv:1303.7215 [hep-ph].

\bibitem{Ball:2013gsa}
  R.~D.~Ball {\it et al.}  [ The NNPDF Collaboration],
  arXiv:1303.1189 [hep-ph].

\bibitem{Thorne:2012az}
  R. Thorne,
  Phys.\ Rev.\ D {\bf 86} (2012) 074017
  [arXiv:1201.6180 [hep-ph]].

\bibitem{Forte:2010ta}
  S.~Forte, E.~Laenen, P.~Nason and J.~Rojo,
  Nucl.\ Phys.\ B {\bf 834} (2010) 116
  [arXiv:1001.2312 [hep-ph]].

\bibitem{Alekhin:2012ig}
  S.~Alekhin, J.~Blumlein and S.~Moch,
  Phys.\ Rev.\ D {\bf 86} (2012) 054009
  [arXiv:1202.2281 [hep-ph]].

\bibitem{Gleisberg:2008ta} 
  T.~Gleisberg, S.~.Hoeche, F.~Krauss, M.~Schonherr, S.~Schumann, F.~Siegert and J.~Winter,
  JHEP {\bf 0902}, 007 (2009)
  [arXiv:0811.4622 [hep-ph]].

\bibitem{Frederix:2011ss} 
  R.~Frederix, S.~Frixione, V.~Hirschi, F.~Maltoni, R.~Pittau and P.~Torrielli,
  JHEP {\bf 1202}, 099 (2012)
  [arXiv:1110.4738 [hep-ph]].

\end{thebibliography}
\end{document}